# AI-guided digital intervention with physiological monitoring reduces intrusive memories after experimental trauma

## Authors
Megan T. deBettencourt[1*], Sruthi Sakthivel[1], Emily A. Holmes[2], Mark Chevillet[1]

## Affiliations
1. Ruby Neurotech, Redwood City, CA, USA
2. Department of Women's and Children's Health, Uppsala University, Uppsala, Sweden

\* Corresponding author, megan@ruby-neurotech.com

**Keywords:** Generative AI, large-language models, neurotechnology, pupillometry, PTSD

## Abstract

Trauma prevalence is vast globally. Evidence-based digital treatments can help, but most require human guidance. Human guides provide tailored instructions and responsiveness to internal cognitive states, but limit scalability. Can generative AI and neurotechnology provide a scalable alternative? Here we test ANTIDOTE, combining AI guidance and pupillometry to automatically deliver and monitor an evidence-based digital treatment, specifically the Imagery Competing Task Intervention (ICTI), to reduce intrusive memories after psychological trauma. One hundred healthy volunteers were exposed to videos of traumatic events and randomly assigned to an intervention or active control condition. As predicted, intervention participants reported significantly fewer intrusive memories over the following week. Post-hoc assessment against clinical rubrics confirmed the AI guide delivered the intervention successfully. Additionally, pupil size tracked intervention engagement and predicted symptom reduction, providing a candidate biomarker of intervention effectiveness. These findings open a path toward rigorous AI-guided digital interventions that can scale to trauma prevalence.

## Introduction

Trauma is unfortunately highly prevalent. Around 70% of people globally will experience a traumatic event during their lifetime(Kessler et al. 2017). In the United States alone, post-traumatic stress disorder (PTSD) affects over 13 million adults(Davis et al. 2022). The scale of the problem, however, is accompanied by the practical challenge of scaling treatments after trauma. Current evidence-based, trauma-focused psychological treatments include prolonged exposure, cognitive processing therapy and cognitive behavioural therapy with a trauma focus (CBT-TF)(Bisson and Olff 2021). However these treatments, even when digitized, require multiple sessions with highly trained clinicians, which limits the capacity to meet the enormous need(Ehlers et al. 2023) and there are ongoing concerns about drop out when implemented in routine clinical practice(Wright et al. 2024). Conversely, pharmacological treatments lend themselves to wider distribution, but are not very effective for PTSD(Bisson and Olff 2021). Thus there is a critical need for more scalable, efficient, and effective interventions after trauma.

One appealing target for interventions is intrusive memories of trauma — involuntary, distressing and sensory memories that repeatedly recur. They are a hallmark symptom of post-traumatic stress disorder (PTSD)(American Psychiatric Association 2013). In a nationally representative sample of over 17,000 trauma-exposed adults in the US, 95% of the individuals who met criteria for PTSD reported experiencing intrusive memories(Martalek et al. 2024). Intrusive memories are also prevalent among trauma-exposed individuals who do not meet criteria for a PTSD diagnosis, i.e. have sub-clinical symptoms(Martalek et al. 2024), as well as individuals with other diagnoses such as depression or anxiety(Holmes and Mathews 2010). Without treatment, individuals with sub-clinical PTSD symptoms may face a growing risk of symptom progression(Lalitha Iyadurai et al. 2019). Their widespread prevalence and broad clinical relevance make intrusive memories a prime target for intervention.

Digital therapeutics have emerged as a promising method for deploying mental health treatments at scale(Wang, Lee, and Shin 2023). One recent and novel digital mental health treatment shown to effectively reduce intrusive memories in randomised controlled trials is the Imagery Competing Task Intervention (ICTI)(Kanstrup et al. 2024; Lalitha Iyadurai et al. 2023; Ramineni et al. 2023). The treatment combines two core cognitive components: a brief reactivation of an intrusive memory, followed by a visuospatial task intended to disrupt memory reconsolidation through activating specific cognitive strategies (e.g., mental rotation, planning, and imagery) (Agren et al. 2023). ICTI was initially tested in a laboratory setting in healthy adults using an experimental model of analogue trauma via the trauma film paradigm(James et al. 2016, 2015), and has since been shown to be safe and effective in several clinical trials, including with emergency department patients(L. Iyadurai et al. 2018; Kanstrup et al. 2021) and healthcare workers exposed to secondary trauma(Ramineni et al. 2023; Kanstrup et al. 2024; Lalitha Iyadurai et al. 2023). Despite this promise, ICTI remains limited in scalability due to its dependence on trained human guides, to provide engaging, interactive, and personally tailored verbal instructions as well as to monitor participants' non-verbal responses and engagement of the intended cognitive strategies.

Removing the reliance on a human guide could allow a scalable deployment of a psychological treatment for intrusive memories (e.g., ICTI) that may better meet the needs of trauma-exposed populations globally. Generative AI systems now have sufficient capabilities to instruct participants and assess their comprehension in ways that instructional videos or static text instructions cannot, via interactive and individualized conversations(Maida et al. 2025; Thirunavukarasu et al. 2023). Physiological signals, such as pupillometry (a putative index of cognitive effort)(van der Wel and van Steenbergen 2018; Kahneman and Beatty 1966; Beatty 1982), are now measurable via lightweight and relatively low-cost devices that can be used outside of controlled laboratory settings(Picanço and Tonneau 2018; Wei et al. 2023). This can enable monitoring of internal cognitive states and strategies during the intervention that are otherwise inaccessible via subjective observation or behavior measures alone(Keene et al. 2022; Clewett, Gasser, and Davachi 2020; Joshi and Gold 2020; Konishi et al. 2017; Madsen and Parra 2023). Incorporating these advances in generative AI and physiology measures into ICTI could enable a more scalable solution that can still provide individualized instructions about how to perform the intervention (e.g. explaining how to emphasize planning and mental rotation during gameplay), and observe pupil size during key portions of the protocol (i.e. memory reactivation and gameplay) to infer cognitive effort.

We developed an intelligent neurotech prototype ANTIDOTE (AI-guided Neurotherapy for Traumatic Intrusions in a Digital Therapeutic; Fig. **1**) to implement the evidence-based treatment ICTI. First, we incorporated generative AI to guide participants through the intervention, delivering structured and interactive instructional conversations. Second, we incorporated physiological monitoring to provide insight into participants' cognitive effort through the intervention. The goal was to develop a unified and automated system to encompass the essential roles of instruction and observation played by the human guide. In the current study, we evaluated ANTIDOTE using a widely used experimental model of analogue trauma (i.e., the trauma film paradigm)(James et al. 2016; Varma et al. 2024). Our primary objective was to test whether ANTIDOTE could produce a reduction in the frequency of intrusive memory, compared to a control. We also explored the quality of intervention instruction provided by the AI guides and examined whether physiological signals—specifically pupil size— tracked intervention engagement.

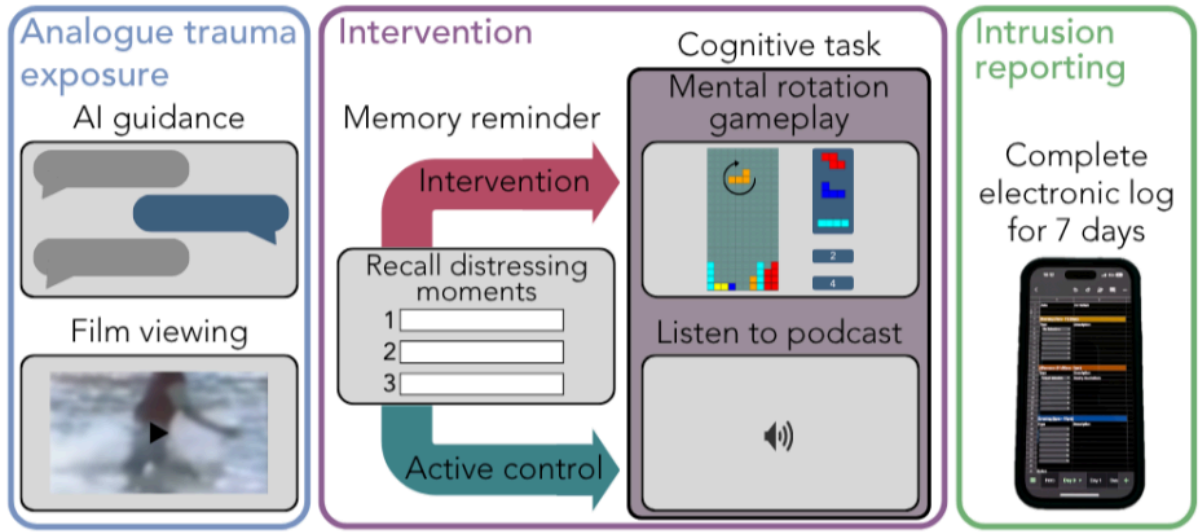

**Figure 1 ANTIDOTE experimental protocol overview.** Participants completed ANTIDOTE, an AI-guided intelligent neurotech prototype to reduce intrusive memory frequency in an experimental model of trauma. Participants were continuously monitored throughout the protocol by physiological monitoring. All participants received instructions throughout the protocol by an AI guide and were exposed to analogue trauma (an 11.5-minute film composed of traumatic video clips). A blurred still from a video is shown for illustrative purposes. After a brief rest period (10 min, not depicted), all participants were given a memory reminder to recall and briefly describe their most distressing moments from the film. Next, participants completed a cognitive task (15 min) according to their random condition assignment. The intervention group (red) played a visuospatial block puzzle game that emphasized mental rotation during computer gameplay, while the active control group (blue) listened to a podcast discussing classical music. Finally, participants completed intrusion reporting, electronically logging and briefly describing the visual details of any intrusive memories from the film for the following 7 days.

## Results

**Reduction of intrusive memories**
The primary preregistered hypothesis was that participants in the intervention condition would report fewer intrusive memories relative to participants in the active control condition. We obtained the total number of memory intrusions reported in the electronic log from each participant during the week following the experimental session (average number of memory intrusions, $m$=16.31, 95% CIs [12.76, 20.34], n=100). Participants in the intervention condition recorded significantly fewer memory intrusions than participants in the active control condition (*intervention* $m_i$=11.62, [8.42, 15.56], $n_i$=50; *control* $m_c$=21.00, [15.04, 28.02], $n_c$=50; one-tailed *p*=0.007; Cohen's *d*=0.49; Fig. **2a**). These results confirm the preregistered hypothesis and demonstrate the effectiveness of ANTIDOTE in delivering automated psychological intervention to reduce intrusive memories.

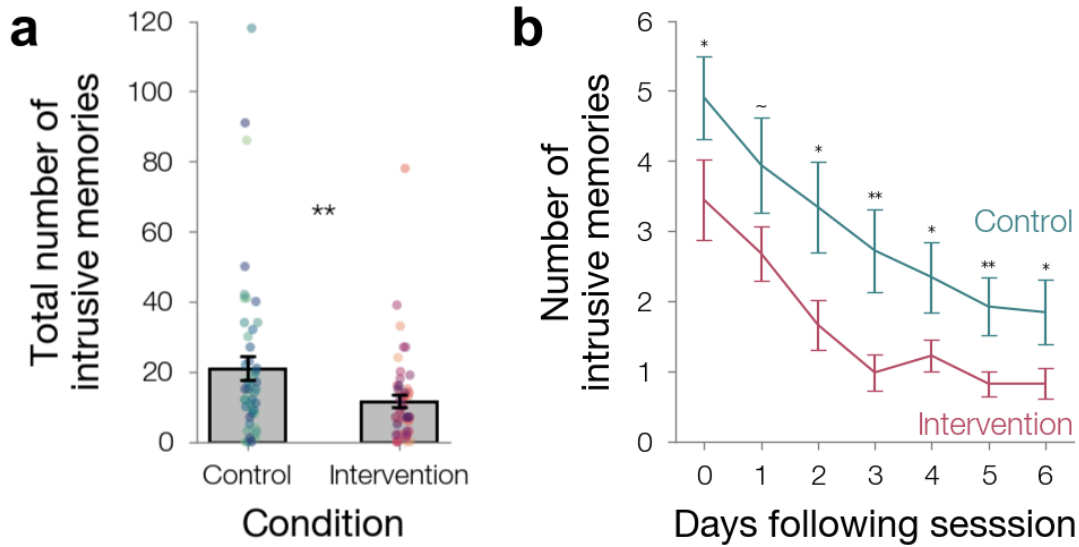

**Figure 2 ANTIDOTE reduces the number of intrusive memories. (a)** Total number of intrusive memories reported in the electronic log per condition. Participants in the intervention condition reported significantly fewer intrusive memories than those in the control condition (** *p*=0.007). The total number of intrusive memories was totaled over the 7-day period following the experimental session. Each dot represents one participant. Bars show group means; error bars indicate standard errors of the mean. **(b)** Time course of intrusive memories reported per day during the 7-day period. Lines indicate group means for the intervention (blue) and control (red) conditions; error bars indicate standard errors of the mean. Statistically significant or trending between-group differences on individual days are marked (** *p*s<0.01, * *p*s<0.05, ~ *p*<0.1).

To understand how the intervention affected the trajectory of intrusive memories over time, we examined the number of intrusive memories reported each day during the 7-day electronic log (Fig. **2b**). We observed reliable differences between conditions on most days throughout the week following the session (one-tailed $p_0$=0.04, $p_1$=0.06, $p_2$=0.01, $p_3$=0.003, $p_4$=0.02, $p_5$=0.007, $p_6$=0.03). To model this trajectory of the change in intrusive memories across days, we fit a mixed-effects linear model with random intercepts for participants (*n*=100; 700 observations). We also observed a significant main effect of condition (β=−1.65, 95% CIs [−2.88, −0.43]; *p*=0.008), consistent with fewer intrusions throughout the time period in the intervention group compared to the active control group. There was also a significant main effect of day (β=−0.51, [−0.61, −0.40]; *p*<0.001), but the interaction between day and condition was not reliable (*β*=0.08, [−0.07,0.23]; *p*=0.30). This pattern suggests that ANTIDOTE exerted a consistent effect across the 7-day period.

**Evaluating AI guidance**

In contrast with previous ICTI studies where the intervention was led by a trained human guide, here an AI guide delivered the instructions through text-based chat conversations with the human participant (see Methods). These human-AI conversations explained each of the key components of the experimental protocol (analogue trauma exposure, intervention condition cognitive task, the concept of intrusive memories, and the rationale and procedure for completing the electronic log) in a structured manner (Fig. **3a**). There was an overall high level of success in AI guidance, as all participants (*n*=100) completed multiple successful conversations with the AI guide.

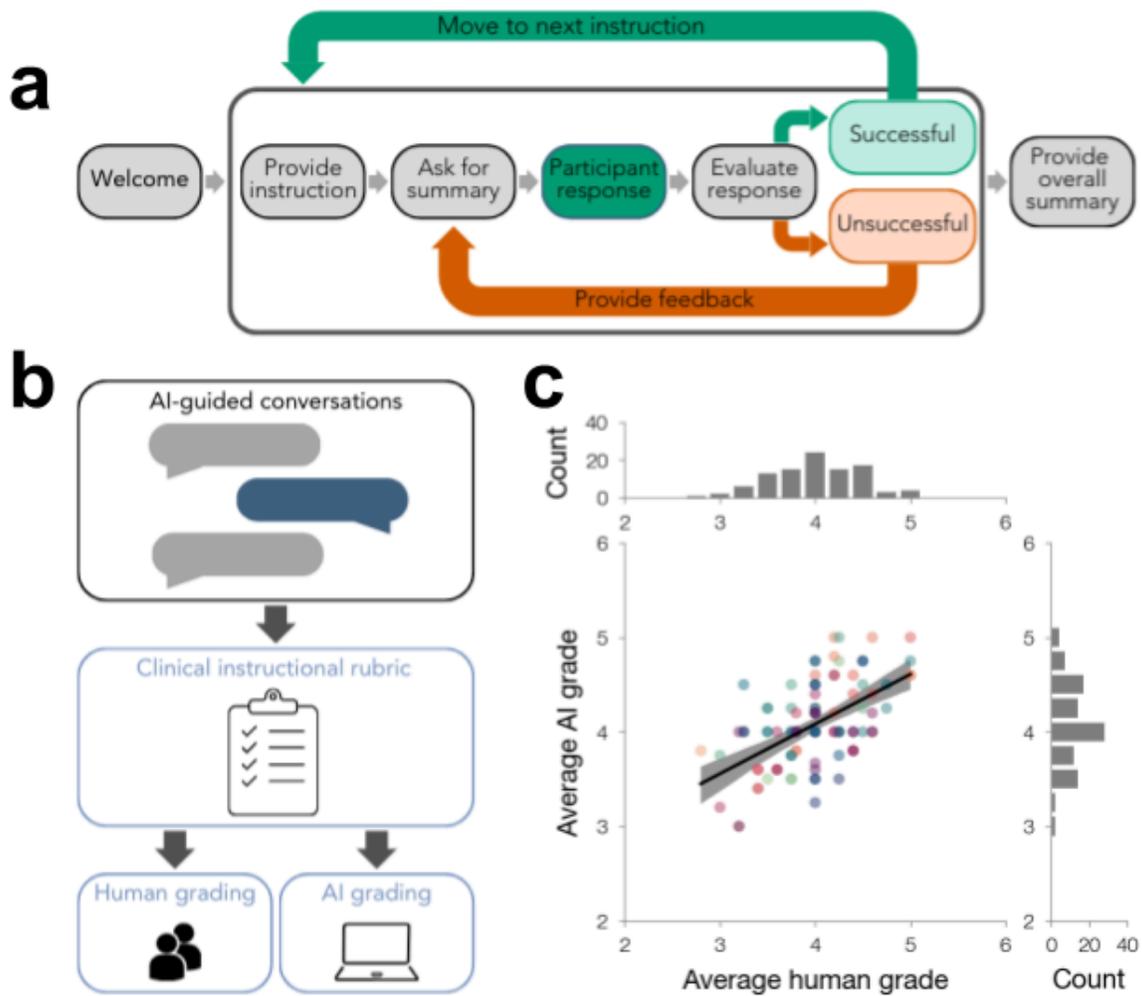

**Figure 3 AI guidance and evaluation. (a)** ANTIDOTE delivered automated instructions throughout key components of the experimental protocol by a series of five human-AI conversations. The AI guide delivered multiple instruction segments, each as a discrete step: the AI presented the instruction, and asked the participant to summarize it, and evaluated the participant's response. If the participant included all key points, the summary for that segment was accepted, and the conversation moved to the next instruction (green). Otherwise, the AI provided corrective feedback and requested a revised summary for that instruction before moving on (red). Upon successful completion of all instruction segments, the AI guide presented a consolidated summary. **(b)** Each AI-guided instruction conversation was evaluated using a rubric previously developed to train human psychology researchers and clinicians to deliver the instructions(Kanstrup et al. 2024). In this study, we used two types of ratings: (i) two human raters independently scored each AI-participant conversation, and (ii) an AI grader assigned scores based on the same criteria. **(c)** Human and AI grading alignment. The human grades (*x*-axis) were strongly correlated with the AI grades (*y*-axis; $p<0.001$). Each dot represents a participant's average score across all conversations. The line depicts the linear fit, and the shaded area reflects 95% CIs. Score distributions for each rater type are projected onto the respective axes as marginal histograms, with bar height indicating the number of participants in that bin.

To better assess the quality of instructional delivery by the AI guide, we conducted a muli-level analysis. These conversations were evaluated in 4 complementary ways: (1) participant survey feedback, (2) human grading of instructional quality, and (3) AI-based grading as a scalable alternative (4) quality control analysis of electronic log entries.

## 1. Participant surveys rate the AI guide highly

First, to understand the participants' experience of using an AI guide, we collected survey data about the experience with the AI guide. In general, participants rated the AI guidance highly (mean rating=4.41 of 5, 95% CIs [4.26, 4.54], *n*=72). To support responsible AI deployment and in alignment with AI safety guidelines, all conversation logs underwent manual post-hoc review for potentially harmful or offensive content, and no such instances were observed.

## 2. Humans grade the human-AI conversations as effective

Second, two human raters manually graded over 400 conversations between the AI guide and the human participant to evaluate instructional quality and participant understanding (Fig. **3b**). They applied a scoring rubric, originally developed for training human therapists to deliver the ICTI intervention(Kanstrup et al. 2024). Each conversation received a consensus integer score between 0 (lowest) and 6 (highest). Overall, the human grading scored the human-AI conversations at a competent level across participants (average score, *s*=4.01, 95% CIs [3.92, 4.10, *n*=100]; Fig. **3c**). Scores were consistent across the five different human-AI conversations: $s_1$=4.01, [3.88, 4.14]; $s_2$=4.00, [3.80, 4.20]; $s_3$=4.07, [3.91, 4.23]; $s_4$=3.72, [3.59, 3.87]; $s_5$=4.27, [4.08, 4.44]. These findings show that the AI guide effectively communicated the instructions.

Furthermore, there was no difference between the score for participants in the intervention (average score $s_i$=4.02, [3.88, 4.16], *n*=50) vs. control conditions (average score $s_c$=4.00, [3.89, 4.12], *n*=50; *p*=0.86). These results indicate that the AI guide reliably delivered instructions with competence and impartiality across conditions.

## 3. AI grading the human-AI conversations

Evaluating the quality of human-AI conversations was a time-intensive task that required manual scoring by trained human graders. We assessed whether an AI grader could produce conversation ratings consistent with human evaluations (Fig. **3b**). We provided the AI grader the same rubric as used by the human graders. The AI grader scored the human-AI conversations at an overall similar level as the human graders (average score, $s$=4.08, 95% CIs [3.99, 4.17], n=100; Fig. **3c**). There was also no reliable difference between the human and AI graders ($p$=0.28, MAE=0.34, RMSE=0.44). Furthermore, grades assigned by the AI were strongly correlated with human scores across participants (Spearman's $\rho$=0.52, $n$=100; $p$<0.001; Fig. **3c**). These findings suggest that AI-based grading offers a scalable alternative for evaluating the fidelity of AI guidance for the intervention.

*4.* <u>Participants successfully complete the electronic log of intrusive memories</u>
We also conducted a quality control analysis of the entries in the electronic logs of intrusive memories to assess whether participants demonstrated a clear understanding of the study definition of the intrusive memory (i.e., image-based descriptions of scenes from the videos watched during the experimental session) and how to successfully complete the log from their conversations with the AI guide. We manually reviewed all 1,631 entries submitted across all participants. Entries with blank descriptions or those that did not meet the study's definition of an intrusive memory were excluded, accounting for 8.03% of the entries. In some cases, single entries captured multiple intrusive memories, resulting in a small increase in the total count (0.06% of the entries). The total number of intrusive memories across participants did not reliably change (Δ=1.15 entries, 95% CIs [0.35, 2.15], $n$=100; $p$=0.68). The fact that so few modifications were made supports that the AI guide successfully conveyed key instructions, enabling participants to understand and complete the electronic log appropriately.

We then conducted a confirmatory analysis, reanalyzing our primary hypothesis of a reduction of intrusive memories in the intervention group versus control. We again observed reliably fewer intrusive memories in the intervention group (intervention: $m_i$=10.70, 95% CIs [7.50, 14.68], $n_i$=50) relative to the control group ($m_c$=19.62, 95% CIs [14.16, 26.14], $n_c$=50; one-tailed $p$=0.006, Cohen's $d$=0.49). This confirms that this result was not driven by data quality issues.

**Imagery competing cognitive task gameplay and pupillometry**
During the cognitive task component of the experimental protocol, participants in the intervention group played a block puzzle game that dynamically varied in difficulty. The game difficulty started at level 1, the slowest and easiest level. When participants successfully cleared a line, the game difficulty increased in a stepwise manner until level 12, the fastest and most difficult level. If the pieces piled up to the top of the game field, the game reset back to level 1. Thus, each participant experienced an individualized trajectory contingent to their game play.

A key aspect of the intervention task is that participants are instructed to engage in mental rotation, planning, and imagery during gameplay. We explored the use of neurophysiological measures, specifically pupil size, a putative signature of cognitive effort(van der Wel and van Steenbergen 2018), to track these internal mental states during gameplay. We compared pupil size versus the 10-minute rest period which occurred after watching the videos. During the intervention cognitive task (i.e., mental rotation gameplay), the average pupil size was larger than during rest (mean difference Δ=0.46, 95% CIs [0.38, 0.54], *n*=47; *p*<0.001; Fig. **4a**). During the control cognitive task (i.e., listening to the podcast), the difference in the average pupil size from rest was trending but not reliable (mean difference Δ=0.05, 95% CIs [0.00, 0.10], *n*=49; *p*=0.07). The interaction between groups was reliable (*p*<0.001).

To more directly link pupil size and cognitive effort, we leveraged the simultaneous dynamics of the game difficulty (Fig. **4b**). For each game piece that fell for every participant, we calculated the difficulty level as well as the mean pupil size. We fit a linear mixed-effects model to examine the relationship between difficulty level and pupil size (*n*=48 participants; 7956 total pieces). Pupil size differences were de-meaned within participants, and both variables (difficulty level and pupil size) were standardized. We included participants as a random effect, with varying intercepts and slopes. There was a reliably positive relationship between the difficulty level and pupil size (*β*=0.26, 95% CIs [0.12, 0.39]; *p*<0.001; Fig. **4c**). That is, pupil size increased with increasing game difficulty.

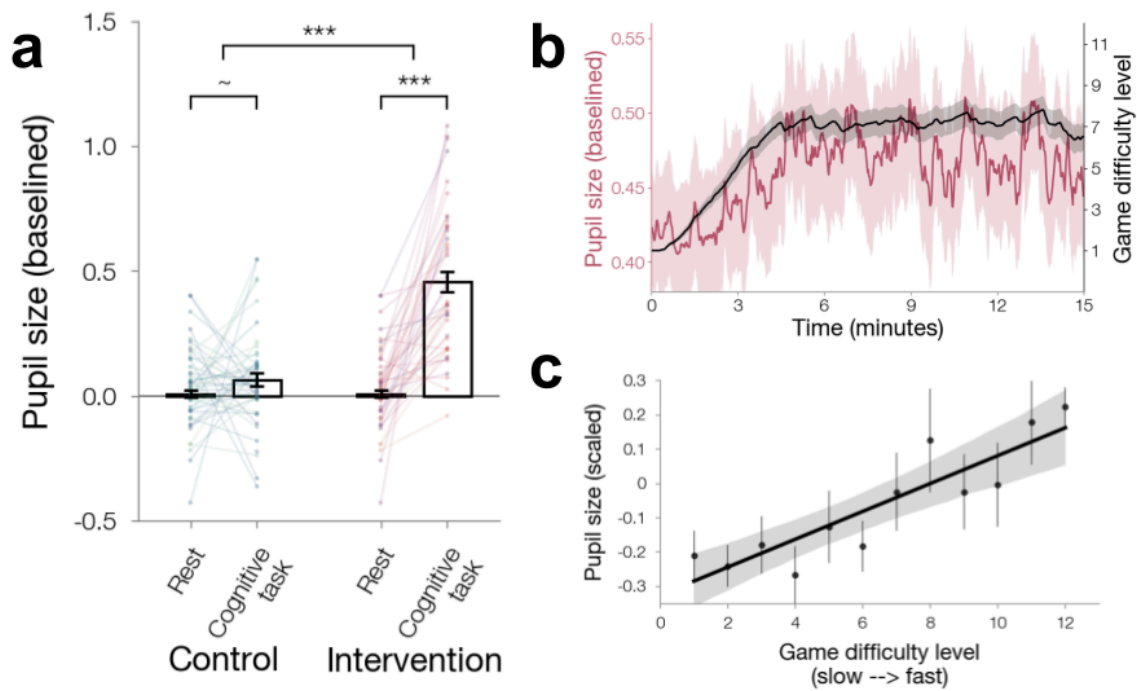

**Figure 4 Pupillometry measures during the cognitive task portion of the intervention.** **(a)** The mean pupil size was calculated for both the intervention and control groups, during the cognitive task and during rest, from a 10-minute period after watching the videos. The pupil size during the cognitive task was reliably greater than during rest for the intervention group (mental rotation gameplay, *** *p*<0.001), and trending but not reliable for the control group (podcast listening, ~ *p*=0.07). The interaction between the intervention and control groups was reliable (*** *p*<0.001). Pupil sizes were baselined to a 3-minute period at the beginning of the experimental protocol. The height of the bar is the population mean, and the error bars show the standard errors of the mean. Each participant is depicted as a dot, and data from the same participant are connected by a line. **(b)** Pupil size and game difficulty dynamically fluctuate over time. Across the 15-minute intervention, the pupil size (red) and game difficulty level (black) varied for each participant in the intervention group. The game difficulty level ranged from 1 (the slowest and starting game speed) to 12 (the fastest and hardest game speed). The lines represent the average trajectory over time for all intervention participants, and the shaded areas are the standard errors of the means. **(c)** Pupil size increases with game difficulty level. A multilevel linear regression was used to model the positive relationship between average pupil size and difficulty level 1–12 ($p<0.001$), accounting for variation across participants in the intervention group. For visualization purposes, the plot depicts an ordinary least squares regression: each dot shows the mean baselined pupil size data (demeaned and scaled within participants) for a given difficulty level, with error bars representing the standard error of the mean and the shaded area representing 95% CIs.

**Memory reminder behavior and pupillometry**

A critical component of ANTIDOTE (and the ICTI intervention) is a memory reminder, when participants are instructed to briefly list their "worst moments" that they remember from the film, prior to the cognitive task. In this paradigm, participants were asked to provide brief descriptions of the key moments that they found most distressing in the videos. Participants listed a variable number of entries (average number of entries #=5.91, 95% CI=[5.48, 6.35], $n$=100). A manual review confirmed that all 591 entries (100%) were related to the video content. There was no significant difference between the number of entries for the control vs. intervention participants ($\#_i$=6.02 [5.40, 6.66]; $\#_c$=5.80, [5.22, 6.38]; $p$=0.66). That is, participants were successful at recalling distressing moments from the films, and the behavioral measure of memory (i.e., number of moments recalled) did not differ between the conditions.

In addition to the memory behavior, we were interested in the internal memory state, which we assessed via pupil size (Fig. **5a**). We predicted that memory reactivation would require cognitive effort, indicated by a larger pupil size. We compared pupil size versus the 10-minute rest period which occurred after watching the videos and prior to the memory reminder. Indeed, pupil size was larger during the memory reminder versus rest, for both participants in the control group (mean difference Δ=0.28, 95% CIs [0.21, 0.36], $n$=49; $p<0.001$) and the intervention group (mean difference Δ=0.24, 95% CIs [0.17, 0.31], $n$=45; $p<0.001$). Consistent with the fact that the memory reminder occurred before the intervention and control groups diverged, there was no significant difference in pupil size between groups ($p$=0.41).

We also examined the pupil dynamics during active memory recall, specifically examining the time between the time at which the reminder screen initially appeared and the time at which the first text entry was submitted. The duration of the entire memory reminder period varied across participants, based on factors including the latency of memory reactivation, number of entries, and typing speed. We aligned the pupil size data from all participants to the onset of the memory reminder screen (Fig. **5b)**. To ensure consistent data length, we truncated each trace to the shortest duration of the memory reminder period (*t*=29 sec) across all participants with available data (*n*=93). At the onset of the memory reminder screen, there was no reliable difference from baseline (*p*=0.79 at t=0 sec). Following a brief initial dip, pupil size rose reliably above baseline (*p*=0.006 at *t*=2.05 sec after the reminder screen appeared) and remained elevated.

Participants provided a variable number of entries during the memory reminder phase (between 1 and 15), modeled after clinical implementations of ICTI for patients with PTSD(Ramineni et al. 2023). Beyond examining the dynamics up until the first entry, we also assessed the effect of subsequent entries. We conducted an exploratory linear mixed-effects model relating entry number (2 and above) to mean pupil size during the entry. The model included all participants with available pupil data from eligible entries (*n*=88 participants, 477 entries total) and incorporated random intercepts and slopes to account for within-subject variability. Both the entry index and pupil size were standardized prior to modeling. As the entry number increased, the pupil size decreased (*β*=−0.18, 95% CI [−0.29, −0.06]; *p*=0.002; Fig. **5c**). According to our interpretation of pupil size as a putative index of cognitive effort, these results suggest that less cognitive effort was expended when reporting later entries.

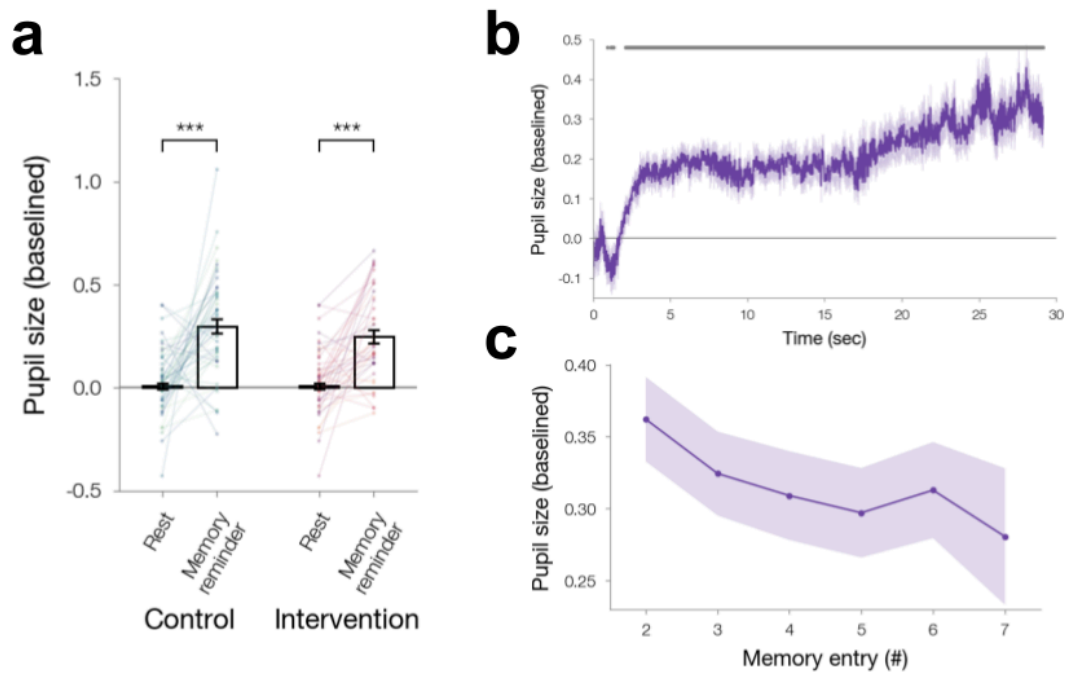

**Figure 5 Pupillometry measures during the memory reminder period of the intervention. (a)** The mean pupil size was calculated for both the intervention and control groups, during the memory reminder and during rest, from a 10-minute period after watching the videos. The pupil size was reliably greater than during rest for both the intervention and control groups (*** *p*s<0.001). There

was no reliable interaction between groups, consistent with the fact that the memory reminder occurred prior to when the intervention and control groups diverged. Pupil sizes were baselined to a 3-minute period at the beginning of the experimental protocol. The height of the bar is the population mean, and the error bars show the standard errors of the mean. Each participant is depicted as a dot, and data from the same participant are connected by a line. **(b)** Pupil size dynamics for the first entry. We analyzed pupil size at the start of memory reactivation, from the onset of the memory reminder screen until the first entry, truncating the window to the shortest duration across all participants ($t$=29 sec). Pupil size was initially not reliably different from baseline; significant time points ($p$s<0.01, uncorrected) are marked by gray along the top of the figure. The purple line shows the mean pupil size relative to baseline, and the shaded area represents the standard error of the mean. **(c)** Pupil size per subsequent memory entries. The purple line shows the mean pupil size for each entry relative to the baseline, the shaded area is the standard error of the mean. For visual clarity, only entries up to 7 are shown, as higher numbers of entries were rare. However, all eligible entries were included in the statistical model. Pupil size was largest for the first entry and declined with increasing entry number ($p$<0.01).

**Physiological predictors of intervention success**

To assess whether our physiological markers of cognitive effort were related to the success of the intervention, we conducted additional exploratory analyses investigating the relationship between pupil size and the number of intrusive memories for two key phases of the experimental session: the cognitive task (either gameplay or listening) and the memory reminder.

*Participants in both the intervention and control groups combined:* First, we investigated the cognitive task phase. We explored whether greater cognitive effort, indexed by larger pupil size during the cognitive tasks (either gameplay or listening), was associated with fewer intrusive memories. Pupil size during the cognitive task was negatively correlated with the number of intrusive memories (Spearman's $\rho$=−0.31, $n$=96; $p$=0.002). This relationship was further quantified by a linear regression of pupil size during the cognitive task predicting the number of intrusive memories ($\beta$=−17.73, [−30.27, −5.19], $n$=96; $p$=0.007; Fig. **6a**). That is, our measure of greater cognitive effort during the cognitive task (i.e., measured during the experimental session) correlated with fewer intrusive memories in the real world over the next week.

Next, we investigated the memory reminder phase. We explored whether the cognitive effort, indexed by the pupil size during the memory reminder phase, was associated with the number of intrusive memories. To test this, we included pupil size measured during the memory reminder period as an additional predictor. Specifically, we fit a linear model (in both intervention and control groups) in which pupil size during both the cognitive task and during the memory reminder period predicted the number of intrusive memories. This allowed us to assess the unique contributions of task engagement and memory reactivation when considered simultaneously. The modeling results revealed that while the influence of the cognitive task remained a significant predictor of the number of intrusive memories (β=−18.51, [−32.89, −4.13], *n*=94; *p*=0.01), the cognitive effort measured during memory reactivation was not reliably predictive (β=0.75, [−18.58, 20.08], *n*=94; *p*=0.94).

*Participants within the intervention group:* The previous analyses examined participants in both the intervention and control groups. We repeated these analyses, restricted to just the participants in the intervention condition. The task effect coefficient replicated the effect found in the full sample, that a larger pupil size during the cognitive task (here gameplay) predicted fewer intrusions (β=−28.41, [−52.33, −4.49], *n*=45; *p*=0.02; Fig. **6b**). Whereas, the coefficient for memory reactivation was reliably positive — a larger pupil size predicted more memory intrusions (β=33.21, [2.14, 64.28], *n*=45; *p*=0.04; Fig. **6b**). That is, both pupil size during the memory reminder and during the mental rotation gameplay task reliably predicted the intervention success, albeit in different directions. This suggests a conceptual model where the ideal approach may be to expend low cognitive effort during memory reactivation, followed by high cognitive effort during the intervention (Fig. **6c**).

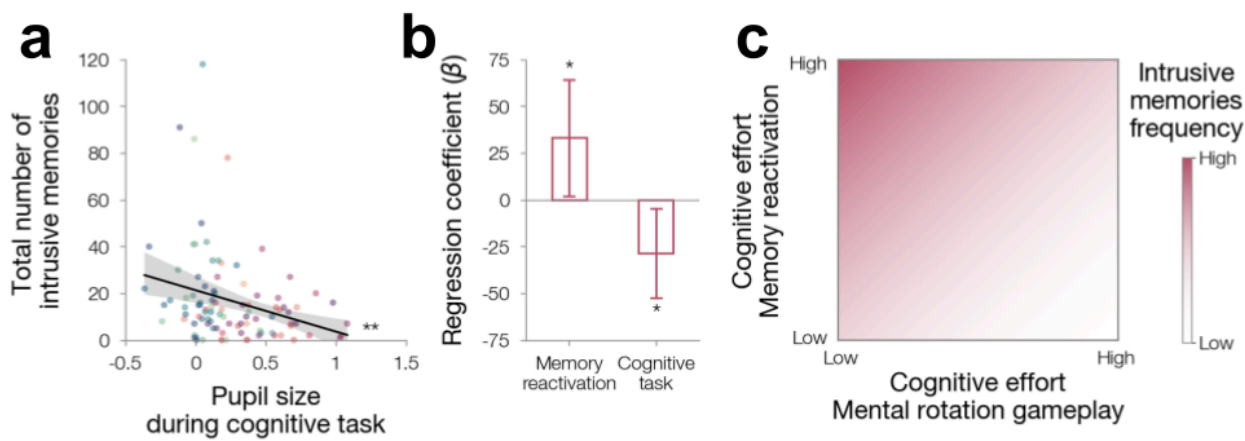

**Figure 6 Physiological predictors of intervention success. (a)** Pupil size during the cognitive task predicts intrusive memories. Mean pupil size during the cognitive task (for both intervention and control groups) was negatively correlated with the number of intrusive memories reported in the 7-day electronic log (** *p*=0.002). Pupil size was baseline-corrected using the pre-task resting period. Each participant is depicted as a dot in a unique color; participants in the intervention group are red and the control group are blue; the line shows the linear fit, and the shaded area reflects the 95% confidence interval. The line depicts the linear fit, and the shaded area is 95% CIs. **(b)** Joint model of memory reactivation and cognitive task effort within the intervention group alone. Both pupil size during memory reactivation and during the cognitive task predicted intrusive memories, but in opposite directions: greater pupil size during the task predicted fewer intrusions, while greater pupil size during reactivation predicted

more intrusions (* *p*s<0.05). The height of the bar is the regression coefficient, the error bars represent the confidence intervals. **(c)** Conceptual model of cognitive effort during ANTIDOTE. To reduce the frequency of intrusive memories, the optimal strategy may be to expend low cognitive effort during memory reactivation, and high cognitive effort during mental rotation gameplay.

**Discussion**

We investigated whether combining advances in generative AI and neurotechnology could allow effective delivery of an existing evidence-based digital mental health treatment in a controlled experimental model of trauma to reduce intrusive memories, and thus enable future scalability. We developed and tested ANTIDOTE, an intelligent neurotech prototype, which combined three key elements: (1) an evidence-based digital treatment for intrusive memories, the Imagery Competing Task Intervention (ICTI), (2) an AI guide to provide interactive instruction and assess participant comprehension, and (3) pupillometry to monitor cognitive effort during key phases of the intervention. We conducted a preregistered randomized controlled experimental study to evaluate whether ANTIDOTE would reduce intrusive memories reported by healthy participants after viewing videos of traumatic events. As hypothesized, participants in the intervention group reported significantly fewer intrusive memories of experimental trauma over the following week compared to an active control. This finding is notable given both conditions were well matched, as each included a memory reminder phase and differed primarily in the type of task that followed (visuospatial versus active control).

We also conducted a series of exploratory analyses to examine how the AI guides and physiological monitoring supported core functions traditionally fulfilled by human guides when administering the digital form of the intervention (i.e., ICTI). The AI guides successfully delivered individualized instruction, as scored on a clinical rubric by both humans and AI, albeit not yet to the same standard of competency of human guides. Additional evidence of instructional effectiveness included favorable participant survey feedback and successful completion of electronic logs containing valid intrusive memory entries. Pupillometry, used to monitor cognitive states during the intervention, provided objective insight into the cognitive effort required during key phases (i.e., memory reminder and gameplay with mental rotation) and predicted intervention outcomes.

*AI Guidance*
There are several possible advantages to the future scalability of digital mental health interventions through the use of generative AI tools(Sharma et al. 2024), specifically large language models (LLMs), to deliver evidence-based experimental protocols. In our study, the AI guide delivered interactive and individualized instructions to participants, providing a consistent and standardized framework for administering the digital intervention. This approach offers key advantages over unblinded human guides, particularly in improving instructional consistency, increasing methodological rigor, and reducing bias. Importantly, the AI guide was unaware of the existence of different treatment groups across participants and was even blinded to condition assignment during different conversations with the same participant.

In addition, the AI guide as an instructional interface neither prompted for nor required disclosure of sensitive personal or health information. This was achieved by careful engineering of the system prompt to constrain the AI guide's function, without therapeutic or diagnostic intent. To further protect participant privacy, we deliberately restricted the AI guide's involvement during the portions of the protocol where participants might be most likely to disclose personally identifiable or sensitive information, i.e., details of the intrusive memory symptom. For both the memory reminder and as participants completed the electronic log of intrusive memories over the following week, the protocol deliberately used static instructions and webforms, rather than AI-guided conversations.

Alongside these advantages, there are also specific limitations and potential future improvements of our use of AI. First, although the guide was not designed to solicit personal disclosures, participants were not explicitly prevented from sharing sensitive information. Future systems could incorporate additional safeguards, such as the use of retrieval augmented generation or automated content moderation, to prevent the risk of unintended disclosures (Inan et al. 2023; Rebedea et al. 2023; Ayyamperumal and Ge 2024).

Second, AI guidance was not used in the podcast listening control task, which may have introduced an imbalance in instructional engagement between conditions. However, conversely, an increase in the verbal and written engagement in the intervention condition may have biased results against the intervention relative to the current control task. Prior work suggests that verbal tasks following trauma exposure do not tend to reduce, and may even increase, the frequency of intrusive memories compared to visuospatial tasks(Holmes et al. 2009). From this perspective, the observed benefit of the ANTIDOTE intervention versus the control may represent a conservative estimate of its potential effectiveness.

Third, while all participants in the current study interacted effectively with the AI guide, future iterations could improve instructional quality and enhance accessibility for diverse populations and individuals with lower digital literacy. Although our grading based on a clinical rubric confirmed the overall competency of the AI guides, they did not reach the highest standard of excellence expected from trained human guides(Kanstrup et al. 2024; Lalitha Iyadurai et al. 2023; Ramineni et al. 2023). Future improvements could reduce the pedantry of the AI guide to increase tolerance for paraphrased input (and discourage parroting or direct copying of the AI's instructions by the participants) and encourage higher-level responses to support comprehension.

*Physiological Monitoring*
A notable innovation of ANTIDOTE is that it incorporated neurotechnology to observe participants during the intervention. In this study, pupil size provided insight into internal cognitive states hypothesized to relate to cognitive effort, such as mental rotation gameplay and trauma memory reactivation, that might otherwise be inaccessible through behavior alone (Keene et al. 2022; Clewett and Murty 2019). If human guidance is not available, physiological monitoring may in the future provide a sensitive metric for treatment compliance. Like motion capture in physical therapy, it can go beyond self-report of intervention completion to confirming that participants executed the intervention as intended and potentially enhancing intervention success(Areias et al. 2024). Pupillometry results also suggested a conceptual model for the various cognitive mechanisms underlying intervention success: low levels of memory reactivation during the memory reminder phase (Detre et al. 2013; Bonsall and Holmes 2023), followed by cognitively effortful gameplay involving mental rotation, mental imagery, and planning (Agren et al. 2023; Kay et al. 2022; Yeung et al. 2025).

While the lighting in the room was held constant during and across sessions, we did not control for low-level visual features on the experimental display that can influence pupil size, such as luminance differences between the tasks used in the intervention and control groups or luminance fluctuations within the mental rotation gameplay(Pan et al. 2024, 2022). This choice was made to preserve consistency with prior studies and to maintain the visual design of the intervention, which may be important for engagement. However, if pupil size primarily reflected visual features rather than cognitive effort, this would likely have weakened rather than strengthened the observed relationships with task difficulty and intervention outcomes—suggesting that the pupillometry signal retained meaningful cognitive information despite any potential luminance confounds.

Future development to further scale the physiological monitoring approach will require more accessible and flexible hardware and software. Although the screen-mounted eye tracker used here was portable, relatively low-cost, and did not require a chin rest, it still limits scalability as it is specialized hardware not typically integrated into standard consumer devices. Emerging methods using standard webcams or smartphone cameras may offer scalable alternatives in the future(Piaggio et al. 2021; Barry et al. 2022). These advances could also enable deployment in more naturalistic settings, while still maintaining measurement fidelity across diverse populations. Finally, although pupil size in this study was monitored in real time, analyses were conducted post hoc. These findings motivate the design of intelligent closed-loop systems that adapt dynamically to optimize cognitive engagement(deBettencourt et al. 2015; Corriveau et al. 2025; Saproo et al. 2016), by adjusting difficulty during the gameplay or the number of memories listed during the memory reminder phase.

*Translational Significance*
Our findings provide proof-of-concept that an AI-guided and physiologically-monitored digital implementation of an evidence-based human-guided digital treatment (ICTI) can reduce the number of intrusive memories in healthy participants after exposure to analogue trauma. The magnitude of this reduction—approximately 45%—closely parallels previous human-guided ICTI laboratory studies using the same trauma film paradigm (51%(James et al. 2015) and 56%(Lau-Zhu, Henson, and Holmes 2021)). While the trauma film paradigm remains a preclinical model, it offers utility for intervention development prior to conducting clinical studies due to the opportunity for strong experimental control(James et al. 2016). Additionally, some interventions developed using this experimental trauma model have been successfully generalized to real-world trauma and intrusive memories(Varma et al. 2024), including human-guided ICTI. Interestingly, even greater reductions in intrusive memories were observed in clinical studies (62% versus active control(L. Iyadurai et al. 2018) and 90% versus waitlist control(Ramineni et al. 2023)). Taken together, this evidence provides a compelling motivation for continued development and clinical translation to test whether future iterations of the approach taken by ANTIDOTE can achieve comparable results in real-world trauma populations.

Beyond the encouraging empirical findings, ANTIDOTE also reflects two emerging directions in digital mental health care. First, the use of LLMs to deliver structured guidance within evidence-based protocols parallels growing efforts to incorporate the use of LLMs in medicine(Thirunavukarasu et al. 2023) and to define and implement the role of digital navigators—human technology coaches increasingly integrated into clinical care teams(Torous et al. 2025). Second, the inclusion of real-time physiological monitoring aligns with increasing interest in integrating objective measures—such as digital phenotyping(Bufano et al. 2023) and digital biomarkers(Coravos, Khozin, and Mandl 2019)—into mental health care and other areas of clinical care where insight into internal state is desirable (e.g. pain)(Fernandez Rojas et al. 2023). While our use of pupillometry was exploratory, it illustrates how physiological signals might eventually support intervention fidelity, cognitive engagement tracking, or personalization of treatment delivery. These trends, though still early in clinical adoption, are beginning to inform implementation models and signal a shift toward more responsive, data-informed approaches in digital mental health(Galatzer-Levy and Onnela 2023).

*Conclusion*
This study establishes initial evidence that a fully automated, AI-guided digital intervention can replicate the intrusion reduction effects of a human-delivered trauma treatment in a controlled experimental model of trauma. By demonstrating that both instruction and engagement monitoring can be delivered without human involvement, ANTIDOTE represents a meaningful step towards scalable, low-cost mental health care. As the use of AI tools and digital phenotyping gain traction in clinical care, approaches like ANTIDOTE that operationalize these concepts in structured, evidence-based interventions could help close the gap between research and real-world impact. Continued development and clinical validation will be critical to determine whether such systems can extend access to effective care for the millions affected by trauma worldwide.

**Methods**

**Experimental protocol overview**

This study developed and evaluated ANTIDOTE, an AI-guided digital neurotech intervention to reduce intrusive memories after exposure to a validated experimental model of trauma(Varma et al. 2024; James et al. 2016). The protocol leveraged an Imagery Competing Task Intervention (ICTI) to reduce intrusive memories, which includes a memory reminder and cognitive task. Using a between-subjects design, participants were randomly assigned to either an intervention or active control condition.

Both the intervention and active control groups underwent highly similar procedures, receiving standardized instructions throughout the protocol by an AI guide while physiological measures were continuously recorded. Both groups watched a film containing traumatic content, followed by a brief rest period, and a memory reminder designed to briefly reactivate participants' memories of the film. The key experimental manipulation occurred during the next phase, the cognitive task. Participants in the intervention condition completed a visuospatial block puzzle game, emphasizing visual imagery and mental rotation. Participants in the active control condition completed an auditory task, listening to a podcast about classical music unlikely to engage visual imagery. Following the experimental session, all participants logged their intrusive memories to the trauma film over the following week using an electronic diary.

**Participants**

One-hundred participants (55 female, 42 male, 3 other/declined to provide sex; mean age = 41.56 years, SD = 14.25, 1 declined to provide age, range 18-65 years) were recruited from the San Francisco Bay Area community via targeted electronic and physical advertisements. This sample size meets and exceeds the preregistered target of 80 ("More than 80 complete datasets may be collected if time and circumstances allow"). Two additional participants started the study but voluntarily disenrolled prior to completion. Inclusion criteria were (a) aged between 18 and 65 years old, (b) English fluency, (c) access to an internet-enabled smartphone or computer, (d) not having previously participated in similar studies, and (e) no self-reported recent or planned stress-inducing or traumatic experiences during the week of study participation. All participants provided their written consent after being informed that the study involved watching emotionally distressing video content and would include both physiological and behavioral measurement. Ethical approval for the study was granted by the Advarra Institutional Review Board (Protocol Reference ID Pro00073795).

**Pre-registration**

The study was pre-registered on the Open Science Framework (OSF) using the template from AsPredicted.org prior to any data collection (https://doi.org/10.17605/OSF.IO/P56JV). The preregistration documented the study design, hypotheses, and planned analyses. The preregistered primary hypothesis was that "*participants receiving the intervention will report fewer intrusive memories relative to participants who receive a control task*". The preregistered primary analysis was to "*compare the total number of intrusive memories reported by participants in the Intervention condition versus those in the Control condition via their entries in the electronic diary over seven days starting from the day of their in-person study session.*" The preregistration also included a series of exploratory analyses, many of which are beyond the scope of this paper. However, we report a subset of exploratory analyses, including whether pupil size differed between groups and whether it predicted the number of intrusive memories reported.

**Condition assignment**
Each participant was randomly and independently allocated to either the intervention (*n*=50) or active control (*n*=50) condition with equal probability. Note, the 50/50 split between conditions occurred by chance, as no stratification or balancing was applied. Randomization was implemented using Python-based random number generation upon each participant's arrival for the experimental session. All instructions were standardized, delivered by the static text within the web-based software platform and the AI guide (see details on how the AI guide was blinded to condition below). This ensured that instructions were consistent across conditions, except for condition-specific instructions related to the cognitive tasks for the intervention or active control groups. The AI was unaware of group assignment or even the existence of multiple groups. Participant blinding was enabled since the consent form stated that participants were to be randomly assigned either to the intervention or active control group, but provided no additional information about either task. The recruiting materials also did not include any depiction of the tasks. Participants were not informed of their group allocation throughout the study.

**AI guide**
A central feature of this study was the use of an AI guide in place of the human guidance used in prior ICTI studies. The AI guide, implemented as a structured chatbot, engaged participants in a structured, multi-turn, multi-phase conversation using only a custom prompt provided to OpenAI's GPT-4 model (i.e., without any fine tuning or retrieval-augmented generation). Each instructional conversation began with a short segment explaining the upcoming task, followed by asking the participant to summarize the instructions in their own words. The AI compared the summary to the original instructions and provided corrective feedback if key points were missing and asked participants to revise their response. Once the summary was deemed complete, the guide proceeded to the next segment.

Participants could ask questions at any point, which the AI would answer before resuming the instructional sequence. Five instructional conversations were interleaved throughout the experimental session, each corresponding to a different phase of the protocol (analogue trauma exposure film viewing, intervention condition cognitive task, the concept and definition of intrusive memories, rationale for intrusive memories log, and procedure for logging intrusive memories). The AI guide was blinded to condition assignment: prompt content was identical across groups, except for the task-specific prompt for the intervention phase. The AI guide retained memory within each conversation to allow for coherent interaction, but had no access to information from previous conversations with the same participant. There was also no access to conversations with other participants. This preserved blinding across all other conversations in the protocol.

**Experiment Protocol**
*Baseline.* At the start of the study, participants completed a 3-minute baseline rest period, during which a fixation cross was displayed on the screen. They were instructed to sit quietly and let their mind wander, but not to close their eyes for an extended period of time or fall asleep. This rest period provided a baseline measure of pupil size across individuals.

*Film viewing.* All participants viewed a compilation of 10 video clips of a distressing nature (approximately 11.5 minutes total duration). The videos included depictions of actual or threatened death and serious injury. The content of the videos was consistent with previous studies using the Trauma Film paradigm(Lau-Zhu, Henson, and Holmes 2019, 2021), including materials such as public service announcements about car accidents, news footage concerning violence towards people or animals, and medical procedures. Before viewing the video clips, participants completed the first instructional conversation with the AI guide, which instructed participants to "immerse" themselves in the scenes, and imagine the events happening to themselves or someone they care about. Participants rated their sadness, depression and hopelessness on a 10-point scale before and after watching the video clips to check for mood change.

*Rest.* Following the video, all participants completed a 10-minute rest period, during which a fixation cross was displayed on the screen. They were instructed to sit quietly and let their mind wander, but not to close their eyes for an extended period of time or fall asleep. This rest period was comparable with previous work(Lau-Zhu, Henson, and Holmes 2019) and here also provided a baseline point of comparison for analyzing physiological signatures of cognitive effort.

*Memory reminder.* All participants were instructed to recall their "worst moments" from the video. These memories have been referred to as hotspots and associated with future intrusive memories(Holmes, Grey, and Young 2005; Grey and Holmes 2008). To enhance AI safety, AI tools were intentionally not used in this task, as it is the one portion within ICTI where participants are asked to disclose potentially sensitive information (i.e. their brief descriptions of intrusive memories). Instead, participants were instructed via static text on the screen to picture the scenes that stood out in their mind and briefly describe the visual details (5-7 words). Participants typed brief descriptions of each scene into a list of entries, so that we could verify whether they had successfully retrieved memories from the videos. Participants could provide a variable number of entries. This design was modeled after clinical implementations of ICTI for patients with PTSD(Ramineni et al. 2023) and intended to enhance ecological validity and enhance the translational relevance of ANTIDOTE.

*Cognitive task.* Participants then completed a 15-minute cognitive task, which differed by condition:

> *Intervention condition.* Participants in the intervention condition completed the imagery competing task. Similar to previous studies, this task involved playing a falling-block puzzle game (a game genre popularized by Tetris). Prior to gameplay, the AI guide instructed participants on how to control the game and the cognitive strategies participants should use. Participants were asked to focus on imagining different placements of each piece rather than maximizing their score. Gameplay lasted 15 minutes, with difficulty (i.e., block drop speed) increasing after each cleared line and resetting when the blocks filled the game field. The software for the game was adapted from publicly available open-source code(Lambert 2019) (CC0 1.0 Universal license).

*Active control condition.* Participants in the control group listened to a podcast containing neutral, non-aversive content. The auditory stimulus was the first 15 minutes of an episode(NPR 2022) of Fresh Air by the US National Public Radio about classical piano, selected so as to not rely heavily on visual imagery nor mental rotation. Instructions for this task were simply to listen to the podcast, and were delivered as static text on the webpage rather than by an instructional conversation with the AI guide.

*Vigilance-intrusion task*. After the cognitive task, all participants (n=100) received instructions from the AI guide explaining the concept of a visual memory intrusion. Following this, most participants (n=98 of 100; 2 excluded due to time constraints) completed a vigilance-intrusion task closely modeled from prior work(Lau-Zhu, Henson, and Holmes 2019). This task provided an opportunity to report intrusive memories during the experimental session by combining a vigilance task (a sustained attention to response task, SART) with concurrent intrusion reporting. Task-specific instructions were provided as static text.

During the task, numbers (0-9) appeared on the screen for 250 ms, followed by a fixation cross for 1500 ms. For 33% of the trials, a blurred still from a video appeared behind the number. In total, participants completed 270 trials.

For the vigilance component (i.e., SART), participants were instructed to press the "j" key in response to every digit except the number 3, which occurred on 10% of the trials. Thus, correctly responding to the number 3 required inhibiting the preprotent response. To report an intrusive memory, participants were instructed to press the "f" key. Unlike previous implementations of this task, participants were not asked to provide written descriptions of each intrusion.

*Intrusive memory reporting.*
At the end of the experimental session, participants completed two final AI-guided instruction conversations: (1) why keeping the intrusive memory log was important to the study, and (2) how to complete the intrusive memory log. For the next seven days, participants logged any intrusive memories of scenes in the videos they experienced by making entries in a Google Sheets spreadsheet using their personal smartphone or computer. The spreadsheet contained multiple tabs: Intro, Example, and then tabs for each day (1-7). The "Intro" tab contained excerpts from the AI guide instruction conversation on intrusion reporting, including the importance of keeping an accurate record for the study, a description of visual intrusions, and instructions for how to complete the electronic log. The Example tab was completed by participants during the AI guide instruction conversation. Daily email reminders linked to the relevant tab. If no intrusions occurred, participants selected "No Intrusions." Otherwise, they selected "Visual Intrusion" and typed a short description. Multiple intrusions were entered as separate rows. Intrusions were counted individually even if repeated. The total number of intrusive memories across all days was used as the primary outcome measure.

**Apparatus**

Participants completed the experimental session alone in a quiet testing room with low ambient lighting. The experimental protocol was presented in the Google Chrome web browser that was displayed on the full screen of an LCD monitor (15.6 inches; 2560x1440 resolution). The experiment was implemented as a JavaScript-based React application using jsPsych (version 7.3.3) alongside custom-built components for the memory reminder and block puzzle game. This React app was hosted on a local server accessed by the client testing machine.

**Physiological data**

Physiological data were collected continuously throughout the experimental session. Data were time-locked to the experimental protocol via WebSocket communication between the front-end React app and the client machine.

Pupil size and eye gaze were recorded using a Tobii Pro Spark eyetracker (60 Hz sampling rate) mounted directly below the LCD monitor. Participants completed a 5-point eyetracker calibration and validation procedure. Due to technical issues, calibration data were not recorded for 2 participants, and eye-tracking data were not recorded for 1 participant. The mean viewing distance during calibration was 68.80 cm (95% CIs [67.34, 70.24], $n$=97). Data analysis of pupillometry data is described in the *Pupillometry Analysis* section below.

Cardiovascular data were recorded using a pulse oximetry ear clip sensor (Nonin Xpod 8000Q2, 75 Hz sampling rate) placed on the participant's left ear lobe. These data were collected to explore heart rate and heart rate variability (HRV) as indicators of cognitive load, analogous to pupil size. However, analysis of these data is beyond the scope of the current paper.

Video recordings were captured using a USB webcam (Logitech Brio, 30 Hz sampling rate) to enable post-hoc assessment of participant compliance with the protocol, since there was no human experimenter in the room with the participant.

**Analysis**

*Total number of intrusive memories*

The primary preregistered hypothesis was that participants in the intervention condition would report fewer intrusive memories than those in the control condition. The primary outcome measure was the total number of intrusive memories recorded by each participant in electronic logs completed during the seven days following the experimental session. We conducted a between-groups comparison of total intrusive memories using a one-tailed permutation test, in line with the directional hypothesis specified in the preregistration.

Following an intention-to-treat approach, we first tested the primary hypothesis using the raw total number of intrusive memory entries from the electronic log from all participants ($n$=100). To assess the robustness of our findings, we conducted two additional confirmatory analysis. First, we evaluated the results after applying quality control procedures to all intrusive memory entries (see *Intrusive memory entry quality control* below). Second, we evaluated the results excluding any participants who deviated from the protocol or were outliers (see *Supplementary Results*). All results remained reliable across all tests of our primary hypothesis.

*Temporal pattern of intrusive memories*

To examine the trajectory of intrusive memories across days, we analyzed between-group differences in the number of the intrusive memories reported per day during the 7-day electronic log. We also modeled these data using a linear mixed-effects model including Day (0-6) and Condition (Intervention or Control) as fixed effects, and participant as a random effect to account for repeated measures.

*Evaluating AI guidance*

All participants completed instructional conversations with the AI guide, which were assessed in 4 ways: (1) self reported surveys from the participants, (2) human grading of the human-AI conversations according to a rubric, (3) AI grading of the human-AI conversations according to the same rubric, and (4) quality control analysis of electronic log entries.

(1) Time permitting, most participants (n=72) completed a brief survey at the end of their experimental session. Participants rated several statements on a 5-point Likert scale (1 = strongly disagree, 5 = strongly agree). Four statements focused on their experience with the AI guide: "The AI chatbot provided instructions that were easy to understand","The AI chatbot was easy to interact with", "The AI chatbot did a good job ensuring I understood the instructions", and "The AI chatbot produced unexpected or inappropriate content.". The final statement was reverse-scored so that, for all items, higher values consistently reflected a more positive experience with the AI guide. Two additional statements assessed their general experience during the session ("I felt physically comfortable throughout the study session.", "The software ran smoothly without any apparent bugs.") and were not included in the analysis of the AI-specific ratings.

(2) Every completed human-AI conversation with each participant was evaluated by human grading. Two human graders applied a rubric closely modeled from one that had previously been used to train human therapists to administer ICTI(Kanstrup et al. 2024; Lalitha Iyadurai et al. 2023; Ramineni et al. 2023). The original rubric was adapted from the revised cognitive therapy scale(Blackburn et al. 2001) and incorporated the Dreyfus system for denoting competence(Dreyfus 1989). The rubric rated the conversations on a seven-point Likert scale, ranging from 0 (absence, no explanation given to participants or explanation is incomprehensible), 1 (major problems), 2 (novice), 3 (advanced beginner), 4 (competent), 5 (proficient), and 6 (excellence, accurate and efficient explanation even in the face of participant difficulties). The two human graders each reviewed the conversation and agreed upon a final consensus score for each conversation. Participant ratings were obtained by averaging the scores from all of their conversations.

(3) We investigated whether an AI model could reliably grade human-AI conversations in a manner consistent with human raters, when provided with the same chat logs and grading rubric. The AI was instructed to evaluate each conversation using the same rubric as the human graders (Open AI, model version gpt-4o). The model was prompted with the grading rubric along with the conversation text, and instructed to assign a numeric score (0-6) along with a brief justification for the rating. Each conversation was evaluated independently, and the prompt remained fixed across all conversations. Participant ratings were obtained by averaging the scores from all of their conversations.

Some participants (*n*=5 of 100) did not complete the fifth and final chat (i.e., the procedure for logging intrusive memories) due to time limitations (n=4) or lack of engagement and falling asleep (n=1). Two of these participants started but did not complete the fourth chat (i.e., the rationale for logging intrusive memories). All incomplete and missing conversations were excluded from scoring by both the human raters and the AI.

(4) We also conducted a post-hoc systematic quality control review of all entries in the electronic log of intrusive memories. Each entry was reviewed in accordance with the study's predefined definition of an intrusive memory, which had three requirements: (1) image-based descriptions of scenes (2) from the videos watched during the experimental session that (3) unintentionally popped into mind. The number of intrusive memories for a participant decreased if the description of the intrusive memory was blank, did not match our definition of intrusive memories (e.g., a verbal rumination), or could not be mapped to a video shown in the experimental session. Conversely, the number of intrusive memories increased if the participant selected "No Intrusions" from the dropdown menu but provided a description of an intrusive memory, if a single log entry could be mapped to multiple intrusions (e.g., "Man shaving and cutting and bleeding x 3") or multiple videos (e.g., "Crushed leg video and elephant"). All adjustments were discussed and agreed upon as a team without consideration of a participant's condition assignment.

*Pupillometry*
Binocular eye tracking was used, and pupil size was averaged across valid samples of left and right eyes. Pupil sizes were baselined to the mean from a 3-minute baseline period at the start of the session. Eye-tracking data were not recorded for one participant due to technical issues, and five others were missing data for specific phases due to either absent synchronization timestamps (from technical or network issues) or if neither eye provided any usable samples, (typically due to tracking loss).

In total, eye-tracking data were recorded for 99 of 100 participants (control: $n_c$=50, intervention: $n_i$=49). We analyzed pupil sizes during four phases of the experiment: baseline, rest, memory reminder, and cognitive task. Pupil data were available from the baseline for 97 participants ($n_c$=49, $n_i$=48), rest for 99 ($n_c$=50, $n_i$=49), memory reminder for 96 ($n_c$=50, $n_i$=46), and cognitive task for 98 ($n_c$=50, $n_i$=48). Statistical analyses comparing baselined pupil size between components (e.g., cognitive task vs. rest) were restricted to participants with valid data in all three components (baseline, cognitive task, and rest), ensuring consistent within-subject comparisons. For the mixed-effects model that related game difficulty to pupil size across pieces, we included participants with valid baselined pupil size data during the intervention cognitive task ($n$=48). For mixed-effects models spanning both phases (cognitive task and memory reminder), we included all participants with valid baselined pupil size data from both phases. Exact sample sizes are reported for each analysis.

**Statistics**
Summary statistics are reported as the mean with 95% Confidence Intervals (CIs). The preregistration stated that *"an independent samples t-test will assess the difference between groups, assuming data normality"* for the primary hypothesis. However, the data used to test the primary hypothesis (i.e., the total number of intrusive memories) violated the assumption of normality, as indicated by the Shapiro-Wilk Test ($p$<0.001) and confirmed through visual inspection. Therefore, statistical tests were conducted using non-parametric permutation tests (100,000 iterations). Despite the non-normality, the results were robust: a parametric test of the primary hypothesis (i.e., independent samples *t*-test) also yielded a consistent and statistically significant pattern of findings (*p*=0.008). In addition, the primary hypothesis was directional in nature ("*participants receiving the intervention will report fewer intrusive memories relative to participants who receive a control task*"). For consistency with the directional primary hypothesis, we used one-tailed *p*-values for its statistical test. In the Results section, we explicitly indicate where one-tailed p-values are reported to maintain transparency. The use of one-tailed *p*-values did not alter the conclusions. The test of the primary analysis in the Results section used the raw total number of intrusive memories from the electronic logs for all participants (*n*=100). We conducted two additional confirmatory analyses of the primary hypothesis: first, following data quality control of the intrusive memory log entries; second after excluding participants with protocol deviations or who were identified as outliers (see Supplementary Results). These adjustments had minimal influence on summary statistics and did not change the outcome of the primary hypothesis.

All other hypotheses were exploratory and non-directional, and therefore two-tailed *p*-values were used for their statistical tests. Correlations were computed using the Spearman rank correlation, which is appropriate for non-parametric data. Analysis of pupillometry data were conducted on all participants with available eye-tracking data from the relevant portions of the study, in order to maximize data inclusion.

We fit linear mixed-effects models using the *statsmodels* package (version 0.14.1), to account for individual differences. We report parametric estimates and associated confidence intervals for the mixed-effects models, as some permutation-based models failed to converge. All analyses were conducted in Python, and all analysis scripts are available in the code repository.

**Data availability**
The data supporting the findings of this study are available in an OSF repository and will be made publicly available upon acceptance. This includes the behavioral data (e.g., intrusive memory counts, reactivation entries, vigilance-intrusion task performance, intervention gameplay metrics), as well as physiological data (e.g., eye-tracking recordings). Due to copyright and privacy concerns, some raw material (such as video stimuli and participant-generated text) cannot be made publicly available but may be available with a material transfers agreement on reasonable request.

**Code availability**
All code used to reproduce the analyses and figures in this study is available in an OSF repository and will be made publicly available upon acceptance.

**Competing Interests**
M.T.dB., S.S., M.C. hold equity in Ruby Neurotech, a company whose interests may be affected by the research reported in this article. E.A.H. is on the Board of Trustees of the MQ Foundation. E.A.H. developed the imagery-competing task intervention for intrusive memories, and know-how in using this over the last 20 years (ANEMONE™ through Afterimagery.AB founded by E.A.H.). E.A.H. receives book royalties from Guildford Press and Oxford University

Press and receives occasional honoraria for conference keynotes and clinical workshops. E.A.H. also receives funding from the Swedish Research Council (2020–00873).

**Acknowledgements**
This work is supported by Wellcome Leap. AI tools assisted with language editing and phrasing suggestions; all content was reviewed and edited by the authors.

**Author Contributions**
All authors conceived the study. S.S. collected the data, M.T.dB. and M.C. supported data collection. M.T.dB. and S.S. analyzed the data and drafted the figures and the manuscript. All authors interpreted the data, reviewed, revised, and approved the final version.

# References

**Bibliography**

Agren, Thomas, Johanna M. Hoppe, Laura Singh, Emily A. Holmes, and Jörgen Rosén. 2023. "The Neural Basis of Tetris Gameplay: Implicating the Role of Visuospatial Processing." *Current Psychology (New Brunswick, N.J.)* 42 (10): 8156–63.

American Psychiatric Association. 2013. *Diagnostic and Statistical Manual of Mental Disorders (DSM-5 (R))*. 5th ed. Arlington, TX: American Psychiatric Association Publishing.

Areias, Anabela C., Dan Doverspike, Daniel F. Brostek, Dora Janela, Michael S. Erwin, John M. Pinter, James R. Ficke, and Fabíola Costa. 2024. "Transforming Veteran Rehabilitation Care: Learnings from a Remote Digital Approach for Musculoskeletal Pain." *Healthcare (Basel, Switzerland)* 12 (15): 1518.

Ayyamperumal, Suriya Ganesh, and Limin Ge. 2024. "Current State of LLM Risks and AI Guardrails." *arXiv [cs.CR]*. arXiv. http://arxiv.org/abs/2406.12934.

Barry, Colin, Jessica De Souza, Yinan Xuan, Jason Holden, Eric Granholm, and Edward Jay Wang. 2022. "At-Home Pupillometry Using Smartphone Facial Identification Cameras." *Proceedings of the SIGCHI Conference on Human Factors in Computing Systems . CHI Conference* 2022 (April). https://doi.org/10.1145/3491102.3502493.

Beatty, Jackson. 1982. "Task-Evoked Pupillary Responses, Processing Load, and the Structure of Processing Resources." *Psychological Bulletin* 91 (2): 276–92.

Bisson, Jonathan I., and Miranda Olff. 2021. "Prevention and Treatment of PTSD: The Current Evidence Base." *European Journal of Psychotraumatology*, January. https://doi.org/10.1080/20008198.2020.1824381.

Blackburn, Ivy-Marie, Ian A. James, Derek L. Milne, Chris Baker, Sally Standart, Anne Garland, and F. Katharina Reichelt. 2001. "The Revised Cognitive Therapy Scale (cts-R): Psychometric Properties." *Behavioural and Cognitive Psychotherapy* 29 (4): 431–46.

Bonsall, Michael B., and Emily A. Holmes. 2023. "Temporal Dynamics of Trauma Memory Persistence." *Journal of the Royal Society, Interface* 20 (203): 20230108.

Bufano, Pasquale, Marco Laurino, Sara Said, Alessandro Tognetti, and Danilo Menicucci. 2023. "Digital Phenotyping for Monitoring Mental Disorders: Systematic Review." *Journal of Medical Internet Research* 25 (1): e46778.

Clewett, David, Camille Gasser, and Lila Davachi. 2020. "Pupil-Linked Arousal Signals Track the Temporal Organization of Events in Memory." *Nature Communications* 11 (1): 4007.

Clewett, David, and Vishnu P. Murty. 2019. "Echoes of Emotions Past: How Neuromodulators Determine What We Recollect." *eNeuro* 6 (2): ENEURO.0108–18.2019.

Coravos, Andrea, Sean Khozin, and Kenneth D. Mandl. 2019. "Developing and Adopting Safe and Effective Digital Biomarkers to Improve Patient Outcomes." *Npj Digital Medicine* 2 (1): 1–5.

Corriveau, Anna, Monica D. Rosenberg, Megan T. deBettencourt, and Megan T. deBettencourt. 2025. "Cognitive Neuroscience of Attention and Memory Dynamics." *PsyArXiv*. https://doi.org/10.31234/osf.io/n7tma_v1.

Davis, Lori L., Jeff Schein, Martin Cloutier, Patrick Gagnon-Sanschagrin, Jessica Maitland, Annette Urganus, Annie Guerin, Patrick Lefebvre, and Christy R. Houle. 2022. "The Economic Burden of Posttraumatic Stress Disorder in the United States from a Societal Perspective." *The Journal of Clinical Psychiatry* 83 (3). https://doi.org/10.4088/JCP.21m14116.

deBettencourt, Megan T., Jonathan D. Cohen, Ray F. Lee, Kenneth A. Norman, and Nicholas B. Turk-Browne. 2015. "Closed-Loop Training of Attention with Real-Time Brain Imaging." *Nature Neuroscience* 18 (3): 470–75.

Detre, Greg J., Annamalai Natarajan, Samuel J. Gershman, and Kenneth A. Norman. 2013. "Moderate Levels of Activation Lead to Forgetting in the Think/no-Think Paradigm." *Neuropsychologia* 51 (12): 2371–88.

Dreyfus, H. L. 1989. *The Dreyfus Model of Skill Acquisition*. Edited by J. Burke, Competency Based Education, and Training. London: Falmer Press.

Ehlers, Anke, Jennifer Wild, Emma Warnock-Parkes, Nick Grey, Hannah Murray, Alice Kerr, Alexander Rozental, et al. 2023. "Therapist-Assisted Online Psychological Therapies Differing in Trauma Focus for Post-Traumatic Stress Disorder (STOP-PTSD): A UK-Based, Single-Blind, Randomised Controlled Trial." *The Lancet. Psychiatry* 10 (8): 608–22.

Fernandez Rojas, Raul, Nicholas Brown, Gordon Waddington, and Roland Goecke. 2023. "A Systematic Review of Neurophysiological Sensing for the Assessment of Acute Pain." *Npj Digital Medicine* 6 (1): 76.

Galatzer-Levy, Isaac R., and Jukka-Pekka Onnela. 2023. "Machine Learning and the Digital Measurement of Psychological Health." *Annual Review of Clinical Psychology* 19 (1): 133–54.

Grey, Nick, and Emily A. Holmes. 2008. "'Hotspots' in Trauma Memories in the Treatment of Post-Traumatic Stress Disorder: A Replication." *Memory (Hove, England)* 16 (7): 788–96.

Holmes, Emily A., Nick Grey, and Kerry A. D. Young. 2005. "Intrusive Images and 'Hotspots' of Trauma Memories in Posttraumatic Stress Disorder: An Exploratory Investigation of Emotions and Cognitive Themes." *Journal of Behavior Therapy and Experimental Psychiatry* 36 (1): 3–17.

Holmes, Emily A., Ella L. James, Thomas Coode-Bate, and Catherine Deeprose. 2009. "Can Playing the Computer Game 'Tetris' Reduce the Build-up of Flashbacks for Trauma? A Proposal from Cognitive Science." *PloS One* 4 (1): e4153.

Holmes, Emily A., and Andrew Mathews. 2010. "Mental Imagery in Emotion and Emotional Disorders." *Clinical Psychology Review* 30 (3): 349–62.

Inan, Hakan, Kartikeya Upasani, Jianfeng Chi, Rashi Rungta, Krithika Iyer, Yuning Mao, Michael Tontchev, et al. 2023. "Llama Guard: LLM-Based Input-Output Safeguard for Human-AI Conversations." *arXiv [cs.CL]*. arXiv. http://arxiv.org/abs/2312.06674.

Iyadurai, Lalitha, Julie Highfield, Marie Kanstrup, Alfred Markham, Varsha Ramineni, Boliang Guo, Thomas Jaki, et al. 2023. "Reducing Intrusive Memories after Trauma via an Imagery-Competing Task Intervention in COVID-19 Intensive Care Staff: A Randomised Controlled Trial." *Translational Psychiatry* 13 (1): 290.

Iyadurai, Lalitha, Renée M. Visser, Alex Lau-Zhu, Kate Porcheret, Antje Horsch, Emily A. Holmes, and Ella L. James. 2019. "Intrusive Memories of Trauma: A Target for Research Bridging Cognitive Science and Its Clinical Application." *Clinical Psychology Review* 69 (April):67–82.

Iyadurai, L., S. E. Blackwell, R. Meiser-Stedman, P. C. Watson, M. B. Bonsall, J. R. Geddes, A. C. Nobre, and E. A. Holmes. 2018. "Preventing Intrusive Memories after Trauma via a Brief Intervention Involving Tetris Computer Game Play in the Emergency Department: A Proof-of-Concept Randomized Controlled Trial." *Molecular Psychiatry* 23 (3): 674–82.

James, Ella L., Michael B. Bonsall, Laura Hoppitt, Elizabeth M. Tunbridge, John R. Geddes, Amy L. Milton, and Emily A. Holmes. 2015. "Computer Game Play Reduces Intrusive Memories of Experimental Trauma via Reconsolidation-Update Mechanisms." *Psychological Science* 26 (8): 1201–15.

James, Ella L., Alex Lau-Zhu, Ian A. Clark, Renée M. Visser, Muriel A. Hagenaars, and Emily A. Holmes. 2016. "The Trauma Film Paradigm as an Experimental Psychopathology Model of Psychological Trauma: Intrusive Memories and beyond." *Clinical Psychology Review* 47 (July):106–42.

Joshi, Siddhartha, and Joshua I. Gold. 2020. "Pupil Size as a Window on Neural Substrates of Cognition." *Trends in Cognitive Sciences* 24 (6): 466–80.

Kahneman, D., and J. Beatty. 1966. "Pupil Diameter and Load on Memory." *Science (New York, N.Y.)* 154 (3756): 1583–85.

Kanstrup, Marie, Laura Singh, Katarina E. Göransson, Julia Widoff, Rod S. Taylor, Beau Gamble, Lalitha Iyadurai, Michelle L. Moulds, and Emily A. Holmes. 2021. "Reducing Intrusive Memories after Trauma via a Brief Cognitive Task Intervention in the Hospital Emergency Department: An Exploratory Pilot Randomised Controlled Trial." *Translational Psychiatry* 11 (1): 30.

Kanstrup, Marie, Laura Singh, Elisabeth Johanna Leehr, Katarina E. Göransson, Sara Ahmed Pihlgren, Lalitha Iyadurai, Oili Dahl, et al. 2024. "A Guided Single Session Intervention to Reduce Intrusive Memories of Work-Related Trauma: A Randomised Controlled Trial with Healthcare Workers in the COVID-19 Pandemic." *BMC Medicine* 22 (1): 403.

Kay, Lachlan, Rebecca Keogh, Thomas Andrillon, and Joel Pearson. 2022. "The Pupillary Light Response as a Physiological Index of Aphantasia, Sensory and Phenomenological Imagery Strength." *eLife* 11 (March). https://doi.org/10.7554/eLife.72484.

Keene, Paul A., Megan T. deBettencourt, Edward Awh, and Edward K. Vogel. 2022. "Pupillometry Signatures of Sustained Attention and Working Memory." *Attention, Perception & Psychophysics* 84 (8): 2472–82.

Kessler, Ronald C., Sergio Aguilar-Gaxiola, Jordi Alonso, Corina Benjet, Evelyn J. Bromet, Graça Cardoso, Louisa Degenhardt, et al. 2017. "Trauma and PTSD in the WHO World Mental Health Surveys." *European Journal of Psychotraumatology* 8 (sup5): 1353383.

Konishi, Mahiko, Kevin Brown, Luca Battaglini, and Jonathan Smallwood. 2017. "When Attention Wanders: Pupillometric Signatures of Fluctuations in External Attention." *Cognition* 168 (November):16–26.

Lambert, Steven. 2019. "Basic Tetris HTML and JavaScript Game." Github. https://gist.github.com/straker/3c98304f8a6a9174efd8292800891ea1.

Lau-Zhu, Alex, Richard N. Henson, and Emily A. Holmes. 2019. "Intrusive Memories and Voluntary Memory of a Trauma Film: Differential Effects of a Cognitive Interference Task after Encoding." *Journal of Experimental Psychology. General* 148 (12): 2154–80.

Lau-Zhu, Alex, Richard N. Henson, and Emily A. Holmes. 2021. "Selectively Interfering with Intrusive but Not Voluntary Memories of a Trauma Film: Accounting for the Role of Associative Memory." *Clinical Psychological Science* 9 (6): 1128–43.

Madsen, Jens, and Lucas C. Parra. 2023. "Narratives Engage Brain and Body: Bidirectional Interactions during Natural Story Listening." *Neuroscience*. bioRxiv. https://www.biorxiv.org/content/10.1101/2023.01.31.526511v2.full.

Maida, Marcello, Daryl Ramai, Yuichi Mori, Mário Dinis-Ribeiro, Antonio Facciorusso, Cesare Hassan, and the AI-CORE (Artificial Intelligence COlorectal cancer Research) Working Group. 2025. "The Role of Generative Language Systems in Increasing Patient Awareness of Colon Cancer Screening." *Endoscopy* 57 (3): 262–68.

Martalek, Alexandra, Caroline Dubertret, Thomas Fovet, Yann Le Strat, and Sarah Tebeka. 2024. "Distressing Memories: A Continuum from Wellness to PTSD." *Journal of Affective Disorders* 363 (October):198–205.

NPR. 2022. *Classical Pianist Jeremy Denk*. National Public Radio. https://www.npr.org/2022/03/21/1087860647/classical-pianist-jeremy-denk.

Pan, Jasmine, Michaela Klímová, Joseph T. McGuire, and Sam Ling. 2022. "Arousal-Based Pupil Modulation Is Dictated by Luminance." *Scientific Reports* 12 (1): 1390.

Pan, Jasmine, Xuelin Sun, Edison Park, Marine Kaufmann, Michaela Klimova, Joseph T. McGuire, and Sam Ling. 2024. "The Effects of Emotional Arousal on Pupil Size Depend on Luminance." *Scientific Reports* 14 (1): 21895.

Piaggio, Davide, Georgy Namm, Paolo Melillo, Francesca Simonelli, Ernesto Iadanza, and Leandro Pecchia. 2021. "Pupillometry via Smartphone for Low-Resource Settings." *Biocybernetics and Biomedical Engineering* 41 (3): 891–902.

Picanço, Carlos R., and François Tonneau. 2018. "A Low-Cost Platform for Eye-Tracking Research: Using Pupil© in Behavior Analysis." *Journal of the Experimental Analysis of Behavior* 110 (2): 157–70.

Ramineni, Varsha, Philip Millroth, Lalitha Iyadurai, Thomas Jaki, Jonathan Kingslake, Julie Highfield, Charlotte Summers, Michael B. Bonsall, and Emily A. Holmes. 2023. "Treating Intrusive Memories after Trauma in Healthcare Workers: A Bayesian Adaptive Randomised Trial Developing an Imagery-Competing Task Intervention." *Molecular Psychiatry* 28 (7): 2985–94.

Rebedea, Traian, Razvan Dinu, Makesh Sreedhar, Christopher Parisien, and Jonathan Cohen. 2023. "NeMo Guardrails: A Toolkit for Controllable and Safe LLM Applications with Programmable Rails." *arXiv [cs.CL]*. arXiv. http://arxiv.org/abs/2310.10501.

Saproo, Sameer, Victor Shih, David C. Jangraw, and Paul Sajda. 2016. "Neural Mechanisms Underlying Catastrophic Failure in Human–machine Interaction during Aerial Navigation." *Journal of Neural Engineering* 13 (6): 066005.

Sharma, Ashish, Kevin Rushton, Inna Wanyin Lin, Theresa Nguyen, and Tim Althoff. 2024. "Facilitating Self-Guided Mental Health Interventions through Human-Language Model Interaction: A Case Study of Cognitive Restructuring." In *Proceedings of the CHI Conference on Human Factors in Computing Systems*, 21:1–29. New York, NY, USA: ACM.

Thirunavukarasu, Arun James, Darren Shu Jeng Ting, Kabilan Elangovan, Laura Gutierrez, Ting Fang Tan, and Daniel Shu Wei Ting. 2023. "Large Language Models in Medicine." *Nature Medicine* 29 (8): 1930–40.

Torous, John, Jake Linardon, Simon B. Goldberg, Shufang Sun, Imogen Bell, Jennifer Nicholas, Lamiece Hassan, Yining Hua, Alyssa Milton, and Joseph Firth. 2025. "The Evolving Field of Digital Mental Health: Current Evidence and Implementation Issues for Smartphone Apps, Generative Artificial Intelligence, and Virtual Reality." *World Psychiatry: Official Journal of the World Psychiatric Association (WPA)* 24 (2): 156–74.

Varma, Mohith M., Shengzi Zeng, Laura Singh, Emily A. Holmes, Jingyun Huang, Man Hey Chiu, and Xiaoqing Hu. 2024. "A Systematic Review and Meta-Analysis of Experimental Methods for Modulating Intrusive Memories Following Lab-Analogue Trauma Exposure in Non-Clinical Populations." *Nature Human Behaviour* 8 (10): 1968–87.

Wang, Changwon, Chungkeun Lee, and Hangsik Shin. 2023. "Digital Therapeutics from Bench to Bedside." *Npj Digital Medicine* 6 (1): 38.

Wei, Wanni, Qing Xue, Xiaonan Yang, Hongjiang Du, Yahui Wang, and Qinglong Tang. 2023. "Assessing the Cognitive Load Arising from in-Vehicle Infotainment Systems Using Pupil Diameter." In *Lecture Notes in Computer Science*, 440–50. Lecture Notes in Computer Science. Cham: Springer Nature Switzerland.

Wel, Pauline van der, and Henk van Steenbergen. 2018. "Pupil Dilation as an Index of Effort in Cognitive Control Tasks: A Review." *Psychonomic Bulletin & Review* 25 (6): 2005–15.

Wright, Simonne, Eirini Karyotaki, Pim Cuijpers, Jonathan Bisson, Davide Papola, Anke B. Witteveen, Sudie E. Back, et al. 2024. "Predictors of Study Dropout in Cognitive-Behavioural Therapy with a Trauma Focus for Post-Traumatic Stress Disorder in Adults: An Individual Participant Data Meta-Analysis." *BMJ Mental Health* 27 (1): e301159.

Yeung, Ryan C., H. Moriah Sokolowski, Carina L. Fan, Myra A. Fernandes, and Brian Levine. 2025. "The Curse of Imagery: Trait Object and Spatial Imagery Differentially Relate to Symptoms of Posttraumatic Stress Disorder." *Clinical Psychological Science*, February. https://doi.org/10.1177/21677026251315118.

**Supplementary Results**

**Mood induction**
To validate that watching the film induced a negative mood, we assessed changes in mood before and after viewing the trauma film ($n$=99, 1 participant's data not recorded due to a technical error). We calculated the increase in sadness (s=4.59 95% CIs [4.11, 5.06], $p$<0.001), depression (d=2.60, 95% CIs [2.16, 3.05], $p$<0.001), and hopelessness (h=2.49, 95% CIs [2.03, 2.98], $p$<0.001). These results confirmed that the film successfully induced a negative mood.

**Robustness of intrusive memory reduction**
The test of the primary hypothesis in the main results used an intention-to-treat analysis, i.e., the full sample. The only participants not included in the analysis were participants who did not complete the study due to voluntary disenrollment ($n$=2, see Methods). We then conducted a series of quality control analyses, excluding participants based on predefined criteria following the preregistration. As noted in the *Outliers and Exclusions* section of the preregistration, "*All exclusion decisions were reviewed and agreed by two or more study members and documented.*"

First, we excluded participants whose "*data were incomplete or the subject was unable to comply with study procedures*" ($n$=11 of 100 participants). The excluded participants had either experienced technical issues during the study ($n$=3), fallen asleep ($n$=2), or received in-person instruction by an experimenter in place of AI instruction due to time limitations ($n$=2), lack of engagement ($n$=2) or lack of comprehension ($n$=2). The reduction of intrusive memories for the intervention vs. control participants remained reliable after these participant exclusions ($m_i$=12.09, 95% CIs [8.55, 16.48], $n$=44 of 50; $m_c$=20.13, 95% CIs [14.36, 27.11], $n$=45 of 50; one-tailed $p$=0.02).

Next, we conducted participant-level outlier exclusion based on behavioral performance. The preregistration mistakenly specified a threshold of "*3 standard errors of the mean*". Applying this overly stringent threshold would have excluded 20 participants (11 control, 9 intervention). Despite this, the group difference remained reliable after outlier exclusions ($m_i$=7.22, 95% CIs [5.63, 8.83], $n$=41 of 50; $m_c$=11.31, 95% CIs [8.90, 13.77], $n$=39 of 50; one-tailed $p$=0.004).

However, this criterion reflected a typographical error in the preregistration. The intended threshold was "3 standard deviations from the mean", not standard errors. Applying the corrected threshold resulted in the exclusion of only 2 participants. The group difference remained reliable after these outlier exclusions ($m_i$=10.27, 95% CIs [7.90, 12.84], $n$=49 of 50; $m_c$=19.02, 95% CIs [14.06, 24.69], $n$=49 of 50; one-tailed $p$=0.002).

In sum, the observed reduction in intrusive memories was robust to exclusion decisions, including preregistered protocol exclusions and outlier handling.

**Comparing in-person and remote assessments of intrusions**
After completing AI-guided ICTI, participants ($n$=98 of 100) completed a vigilance-intrusion task designed to assess the occurrence of intrusive memories at the end of the experimental session. Participants performed the vigilance portion of the task accurately (average accuracy=83.31%, 95% CIs [81.04, 85.46]), and also reported a substantial number of intrusions during the task (average # intrusions, #=49.60, [43.52, 55.78]). Supporting the validity of this task as a measure of susceptibility to intrusive memories, the number of intrusions reported during the vigilance-intrusion task correlated with the number of intrusive memories recorded in the week-long electronic log (Spearman's $\rho$=0.23, $p$=0.02, $n$=98).

However, there was no reliable difference in the number of intrusions reported during the vigilance-intrusion tasks between the intervention and control groups ($\#_i$=51.10 [41.90, 60.46], $n$=40; $\#_c$=48.04, [40.08, 66.46], $n$=48; $p$=0.63). This differs from prior work that has observed a group difference in the vigilance-intrusion tasks(Lau-Zhu, Henson, and Holmes 2019). Future work could investigate this by reducing the differences between prior work, incorporating an AI guide to provide interactive task instructions and collecting brief descriptions of the intrusions to ensure their accuracy. Remote diaries, such as the electronic log, are the clinical gold standard for assessing intrusive memories, and remain the best way to assess the success of the intervention.